\documentclass[prl,twocolumn,aps,a4paper,superscriptaddress,showpacs]{revtex4}
\usepackage{amsmath}
\usepackage[dvips]{graphicx}
\begin{document}

\newcommand\be{\begin{equation}}
\newcommand\ee{\end{equation}}
\newcommand\bea{\begin{eqnarray}}
\newcommand\eea{\end{eqnarray}}
\newcommand\bseq{\begin{subequations}} %solo con amsmath
\newcommand\eseq{\end{subequations}}
\newcommand\bcas{\begin{cases}}
\newcommand\ecas{\end{cases}}
\newcommand{\p}{\partial}
\newcommand{\f}{\frac}

\title{Evolutionary Quantum Dynamics of a Generic Universe}

%\date{\today}
%%%%%%%%%%%%%%%%%%%%%%%%%%%%%%%%
% \email{}
%\affiliation{}
%\superscriptaddress{}
\author{Marco Valerio Battisti}
\email{battisti@icra.it}
\affiliation{ICRA - International Center for Relativistic Astrophysics}
\affiliation{Dipartimento di Fisica (G9), Universit\`a di Roma ``La Sapienza'' P.le A. Moro 5, 00185 Roma, Italy}
\author{Giovanni Montani}
\email{montani@icra.it} 
\affiliation{ICRA - International Center for Relativistic Astrophysics}
\affiliation{Dipartimento di Fisica (G9), Universit\`a di Roma ``La Sapienza'' P.le A. Moro 5, 00185 Roma, Italy}
\affiliation{ENEA C.R. Frascati (U.T.S. Fusione), Via Enrico Fermi 45, 00044 Frascati, Roma, Italy}

\today

\begin{abstract}
The implications of an Evolutionary Quantum Gravity are
addressed in view of formulating a new dark matter candidate.
We consider a Schr\"odinger dynamics for the gravitational field
associated to a generic cosmological model and then we solve
the corresponding eigenvalues problem,
inferring its phenomenological
issue for the actual Universe.
The spectrum of the super-Hamiltonian is determined
including a free inflaton field, the ultrarelativistic
thermal bath and a
perfect gas into the dynamics. We show that, when a Planckian cut-off is
imposed in the theory and the classical limit of the ground state
is taken, then a dark matter contribution can not arise
because its critical parameter $\Omega _{dm}$
is negligible today when the appropriate 
cosmological implementation of the model is provided.
Thus, we show that, from a phenomenological point of view, 
an Evolutionary Quantum Cosmology overlaps the Wheeler-DeWitt
approach and therefore it can be inferred as appropriate to 
describe early stages of the Universe without significant traces
on the later evolution. Finally, we provide indications that
the horizon paradox can be solved in the Planck era by the
morphology of the Universe wave function. 
\end{abstract}


\maketitle

\section{Introduction}

A long standing problem in Quantum Gravity is the
so-called {\it frozen formalism} characterizing the canonical
approach, i.e. the absence of a time evolution for the wavefunctional
\cite{DeW67,Ku81} (see also \cite{Ish92,ShSia,ShSib} and references therein). In a recent
approach \cite{Mo02,MeMo04a}, it has been proposed that such feature
disappears as soon as the impossibility of a physical slicing
without frame fixing is recognized for a quantum space-time.
Since the existence of a timelike direction is ensured on a
quantum level only by including a fluid into the dynamics
(its lightcone structure has to survive), then an equivalence between
time and a frame of reference can be established \cite{MeMo04b} (see also \cite{KuTo,KuBr} and \cite{Ro91a,Ro91b}). 

In the light of these results, here we address a Schr\"odinger
Quantum Cosmology relative to a generic inhomogeneous
Universe and discuss the phenomenological implications
of the super-Hamiltonian spectrum in view
of a dark matter candidate. While previous cosmological
implementations of the same quantum dynamics  
\cite{Mo03,CoMo04} were restricted to the isotropic symmetry, here the
real quantum Big-Bang picture is outlined, showing how the energy of the
gravitational and matter fields (vanishing in the
standard Wheeler-DeWitt approach \cite{DeW67}) provides
in the classical limit, a non relativistic relic component.

To reproduce the primordial Universe we include in the
dynamics a free inflaton field, the ultrarelativistic energy
density of the thermal bath and a perfect gas contribution
to account for Planck mass particles.
To develop the phenomenological implications of the spectrum,
we impose a cut-off at the Planck scale and then we show
that the critical parameter of our dark matter candidate
can not fulfill unity in correspondence
to an efficient inflationary scenario;
in fact our dark matter candidate would work only for 
an {\it e-folding} of about $23$, too small for a solution
of the horizons paradox.
In this respect our model offers quantum features of the
dynamics which outline a possible explanation of such a paradox
within the Planck era.

The structure of the letter is the following. In Sec.I we
discuss the Schr\"odinger Quantum Gravity and the picture
resulting from its classical limit; dualism
existing in Quantum Gravity between time
and matter field is stressed. 
In Sec.II the eigenvalue
problem corresponding to the generic inhomogeneous Universe
is formulated and solved in Sec.III. Sec.IV is devoted to discuss
the classical limit of the model, both in the sense of a
WKB approximation, and with respect to the spectrum. 
Finally, in Sec.V we implement the Planckian cut-off and
study the implications on the spectrum cut-off; 
the phenomenological issues in view of a dark matter scenario are
then investigated and concluding remarks outlined. 

\section{I. Schr\"{o}dinger Quantum Gravity}

In this section, in agreement with the point of view
outlined in \cite{KuTo,Mo02,MeMo04a,MeMo04b,Mo03}, we analyze
the implication of a Schr\"{o}dinger formulation
of the quantum dynamics for the gravitational field (A) and
then we establish a dualism between time evolution and matter
fields (B).
       
\vspace{0.6cm}

\subsection{A. Evolutionary Quantum Gravity}

In order to preserve the covariance on space-like
hypersurfaces, we define the state functional $\Psi$
on the Wheeler Superspace of the 3-geometries
$\left\{h_{ij}\right\}$ ($i,j=1,2,3$) (i.e.
it is annihilated by the super-momentum operator
$\hat{H_i}$) and we require the theory to
evolve along the space-time slicing so
that $\Psi=\Psi(t,\left\{h_{ij}\right\})$.

The quantum evolution of the gravitational field is then governed by the smeared Schr\"{o}dinger equation
(we works in units where $\hbar=c=1$, apart from where we refer to the classical limit in which we restore $\hbar$)
\be
i\p_t \Psi=\hat{\mathcal{H}}\Psi\equiv\int_{\Sigma_t^3}d^3x\left(N\hat{H}\right)\Psi
\ee

being $\hat{H}$ the super-Hamiltonian operator, $N$ the lapse function and $\Sigma_t^3$ the one-parameter family of compact boundaryless 3-hypersurfaces which fill the space-time.

Let us now take the following expansion for the wave functional
\be
\Psi=\int D\epsilon\chi(\epsilon,\left\{h_{ij}\right\}) \exp\left\{-i\int_{t_0}^tdt'\int_{\Sigma_t^3}d^3x(N\epsilon)\right\},
\ee

$D\epsilon$ being the Lebesgue measure in the space of the functions $\epsilon(x^i)$. Such an expansion reduces the Schr\"{o}dinger dynamics to an eigenvalues problem of the form
\be\label{eipro}%
\hat{H}\chi=\epsilon\chi,\qquad \hat{H_i}\chi=0,
\ee

which outlines the appearance of a non zero super-Hamiltonian eigenvalue.

In order to reconstruct the classical limit of the above dynamical constraints, we address the limit $\hbar\rightarrow0$ and replace the wave functional $\chi$ by its corresponding zero-order WKB approximation $\chi\sim e^{iS/\hbar}$. Under these restrictions, the eigenvalues problem (\ref{eipro}) reduces to the following classical counterpart
\be
\hat{HJ}S=\epsilon\equiv-2\sqrt h T_{00},\qquad \hat{HJ_i}S=0
\ee

where $\hat{HJ}$ and $\hat{HJ_i}$ denote operators which, acting on the phase $S$, reproduce the super-Hamiltonian and super-momentum Hamilton-Jacobi equations respectively. We see that the classical limit of the adopted Schr\"{o}dinger quantum dynamics is characterized by the appearance of a new matter contribution (associated with the non zero eigenvalue $\epsilon$) whose energy density reads
\be\label{enden}%
\rho\equiv T_{00}=-\f{\epsilon(x^i)}{2\sqrt h},
\ee  

where by $T_{\mu}^{\nu}$ ($\mu,\nu=0,1,2,3$) we refer to the new matter energy-momentum tensor and $h\equiv$~det~$h_{ij}$.

Since the spectrum of the super-Hamiltonian has,
in general, a negative component, we can then infer that,
when the gravitational field is in the ground state,
this matter out-coming in the classical limit has a positive
energy density. The explicit form of (\ref{enden}) is that
of a dust fluid co-moving with the slicing 3-hypersurfaces,
i.e. the field $n^{\mu}$ begin the 4-velocity
normal to the 3-hypersurfaces (in other words,
we deal with an energy-momentum tensor
$T_{\mu\nu}=\rho n_\mu n_\nu$).

This matter contribution has to
emerge in any system which undergoes
a classical limit and therefore it must concern
the history of the Universe.
Indeed, a non-relativistic matter contribution
is observed in the actual Universe and it corresponds to the dark matter component. Since our fluid is non relativistic in the very early stages of the Universe evolution, we are lead to regard it as a good candidate for the so-called {\it cold dark matter} \cite{Ro,Os,Ei,Ru} (for a discussion on the correlation exiting between the observed dark matter and the baryonic one see \cite{Donato} and for a theoretical interpretation of this feature see \cite{Ki06}). In this respect we below investigate the cosmological issue of a Schr\"{o}dinger approach to Quantum Cosmology.

In view of the later analysis, it is worth noting
that in a Sch\"odinger Quantum Gravity, it is possible to
turn the solution space into Hilbert one and therefore a
notion of probability density naturally arises, from the squared modulus of
the wavefunctional. 

\vspace{0.6cm}

\subsection{B. Time-Matter Dualism}

Let us now consider a gravitational system in the presence of
a macroscopic matter source, described by a perfect fluid having a
generic equation of state $p = (\xi - 1 )\rho$\\
($p$ being the pressure and $\xi$ the polytropic index). The energy-momentum
tensor, associated to this system reads
\begin{equation}
T_{\mu \nu } =
\xi \rho u_{\mu }u_{\nu } - (\xi -1)\rho g_{\mu \nu}
\, .
\label{tens}
\end{equation}

To fix the constraints when matter is included in the
dynamics, let us make use of the relations
\be\label{ten2}%
G_{\mu \nu} n^{\mu }n^{\nu } = - \kappa \frac{H}{2\sqrt{h}}
\ee
\be\label{due}%
G_{\mu \nu} n^{\mu }\partial _iy^{\nu } = \kappa \frac{H_i}{2\sqrt{h}}, 
\ee

where $\kappa$ denotes the Einstein constant ($\kappa=8\pi G$) and $\partial _iy^{\mu }$ are the tangent vectors to the
3-hypersurfaces $n_{\mu }\partial _iy^{\mu } = 0$. 
Equations (\ref{ten2}) and (\ref{due}), by (\ref{tens}) and identifying $u_{\mu }$ with $n_{\mu }$
(i.e. the physical space is filled by the fluid),
rewrite
\be
\label{ten3}
\rho = -\frac{H}{2\sqrt{h}}, \qquad H_i = 0;      
\ee

furthermore, we get the equations
\begin{equation}
G_{\mu \nu }\partial _iy^{\mu }\partial _jy^{\nu }
\equiv G_{ij} = \kappa (\xi -1)\rho h_{ij}
\, .
\label{ij}
\end{equation}

We now observe that the conservation law
$T_{\mu ;\nu}^{\nu } = 0$
(semicolon denoting covariant derivative) implies the
following two conditions
\be\label{ten4}%
\xi \left( \rho u^{\mu }\right) _{;\mu } =
(\xi - 1) u^{\mu }\partial _{\mu }\rho 
\ee
\be\label{ten5}%
u^{\nu } u_{\mu ;\nu } = 
\left( 1 - \frac{1}{\xi }\right)
\left( \partial _{\mu } \ln \rho - u_{\mu }u^{\nu }
\partial _{\nu }\ln \rho \right)
\, .
\ee 

If we now adapt the spacetime slicing,
looking the dynamics into the fluid frame
(i.e. $n^{\mu } = \delta _0^{\mu }$), then, by the relation
$n^{\mu } = (1/N, \; -N^i/N)$, we see that the co-moving
constraint implies the synchronous nature of the reference frame.
As it is well-known that a synchronous reference is also a geodesic
one, the right-hand-side of equation 
(\ref{ten5}) must vanish identically and, for a generic
inhomogeneous case, this means to require
$\xi \equiv 1$. Hence, equations (\ref{ten4})
yields $\rho = -\bar{\epsilon }(x^l)/2\sqrt{h}$; 
substituting the last expression into  (\ref{ten3}), we get
the same Hamiltonian constraints associated to the Evolutionary
Quantum Gravity at the point i), as soon as the function
$\bar{\epsilon}$ is turned into the eigenvalue $\epsilon$.
In this respect, we stress that, while
$\bar{\epsilon}$ is positive by definition, the corresponding
eigenvalue can also take negative values because of
the $H$-structure.

Thus, we conclude that a dust fluid is a good choice to realize
a clock in Quantum Gravity, because it induces a non-zero
super-Hamiltonian eigenvalue into the dynamics;
furthermore, for vanishing pressure ($\xi = 1$), the
equations (\ref{ij}) reduces to the right vacuum evolution 
for $h_{ij}$.

\vspace{0.6cm}

The above two points outline, in Quantum Gravity, 
a real dualism
between time evolution and the presence of a dust fluid.
Though the classical limit of a Schr\"odinger Quantum Gravity
always provides (at least in principle) a
\emph{cold dark matter candidate}
according to the discussion A, nevertheless below we
analyze the generic inhomogeneous cosmological solution in the 
view-point B.

The quantum behavior of the Universe expectably concerns
the Planck era and therefore non-relativistic (pressureless)
components are not available as physical clock.
However, Planck mass particles can be reliably inferred and
appropriately described by a perfect gas contribution.
We construct a solution in which such a term is included
and we link the quantum number, associated to its energy
density, to the eigenvalue $\epsilon $.
In doing that, we use the Planck mass particle as a clock
for the quantum dynamics and, at the same level, we induce
by them a non-relativistic matter component into the early
Universe. 

\section{II. The Generic Quantum Universe}

We now apply the Evolutionary Quantum Gravity approach to the dynamics of a generic inhomogeneous Universe, i.e. a model in which any specific symmetry has been removed. There are some reliable indications \cite{KiMo02,BeMo04} that the early stages of the Universe evolution are characterized by such a degree of generality, but in the quantum regime, dealing with the absence of global symmetry, it is required by the indeterminism; in fact on different causal regions the geometry `fluctuates' independently, so preventing global isometries.

A generic cosmological model has to contain four physically independent functions of the spatial coordinates \cite{BKL82} and, in the ADM formalism \cite{KiMo02,BeMo04}, its line element reads     
\begin{equation}\label{metrten}% 
ds^2=N^2dt^2-h_{ij}(dx^i+N^idt)(dx^j+N^jdt)
\end{equation}

with $N^i$ the shift vector and $h_{ij}$ the 3-metric
\begin{equation}
h_{ij}=e^{q_a}\delta_{ad}O^a_b O^d_c \partial_i y^b \partial_j y^c,\ \ \ 
a,b,c,d=1,2,3
\end{equation}

where $q_a$ and $y^a$ are dynamical variables, while $O_b^a(x^i)$ denote three rotation matrices. By virtue of the analysis developed in \cite{BeMo04}, the dynamics of this generic model can be summarized, asymptotically to the Big-Bang, by the following variational principle
\be
\delta S=\delta\int_{\Sigma_t^3\times\Re} dt d^3x(p_a \p_t q^a -NH)
\ee

where adopting Misner-like variables ($R$, $\beta_+$, $\beta_-$) \cite{Mis69a}, the super-Hamiltonian has the structure
\begin{multline}\label{eigenpro}% 
H(x^i)=\kappa \left[-\f{p_R^2}R+\f{1}{R^3}\left(p^2_+ +p^2_-\right)\right]+\f{3}{8\pi}\f{p^2_\phi}{R^3}+
\\-\f{R^3} {4\kappa l^2_{in}} V(\beta_\pm)+\f{\mu^2}R+\f{\sigma^2}{R^2},
\end{multline}

where $l_{in}$ denotes the comoving physical scale of inhomogeneities and the potential term accounts for the spatial curvature of the model:
\be
V(\beta_\pm)=e^{-8\beta_+}+e^{4(\beta_++\sqrt3 \beta_-)}+e^{4(\beta_+-\sqrt3\beta_-)}.
\ee

The $R$ variable describes the isotropic Universe expansion, while $\beta_\pm$ are associated to the space anisotropies. The following three remarks about the above super-Hamiltonian are relevant, as it appears in the {\it long-wavelength approximation} \cite{Mo04}.

i) Since the relation $\p_t y^a=N^i\p_i y^a$ holds \cite{BeMo04}, in a synchronous reference ($N=1$, $N^i=0$) the variables $y^a$ become passive functions and the dynamics reduces, point by point, to the one of a Bianchi IX model \cite{KiMo97,ImMo01}. As show in \cite{BKL82,Ki93,Mo95,BeMo04}, the classical behavior of a generic model is chaotic and, in this representation, each space point stands for a causal region (cosmological horizon).

ii) We included in the dynamics a scalar field $\phi$ in order to account for the inflaton field which lives in the very early Universe. The corresponding potential term has been omitted because negligible at temperatures corresponding to the quantum Universe \cite{Kolb}. Furthermore, it is well know that a minimally coupled (classical) scalar field can suppress Mixmaster oscillations in the approach to the singularity of generic cosmological spacetimes \cite{Be00}.

iii) We add two different contributions, an ultrarelativistic term (associated to $\mu^2(x^i)$) and a perfect gas one (associated to $\sigma^2(x^i)$) \cite{Mac} to the energy of the system. The former term describes the thermal bath of all fundamental particles (whose mass is negligible with respect to the temperature); the latter is due to particles of Planck mass that, in the quantum Universe, can be described as a perfect gas.

The canonical quantization of this model \cite{Mis69b} is reached by replacing the momenta $p_x$ ($x=R,\pm,\phi$) with the operators $\hat{p_x}=-i\p_x$. The corresponding dynamics is summarized by the eigenvalue problem (\ref{eipro}), which here reads (addressing the normal order defined in \cite{Mo02})
\begin{multline}\label{H}%
\hat{H}\chi\equiv\left\{\kappa \left[\p_R\f 1 R\p_R-\f 1 {R^3}\left(\p_+^2+\p_-^2\right)\right]\right.-\f{3}{8\pi}\f1{R^3}\p_\phi^2 +\\ \left.-\f{R^3} {4\kappa l^2_{in}}V(\beta_\pm)+\f{\mu^2}R+\f{\sigma^2}{R^2}\right\}\chi= \epsilon\chi,
\end{multline} 

Since in the Evolutionary Quantum Gravity
$|\Psi|^2$ provides a probability density and in a
stationary state it coincides with
$|\chi|^2$, then 
boundary conditions appropriate to this problem take the form 

\be
\begin{cases}
\chi(R=0,\beta_{\pm},\phi)&<\infty \\
\chi(R\rightarrow\infty,\beta_{\pm},\phi)&=0.
\end{cases}
\ee            

In fact, we investigate the quantum dynamics near the Big-Bang
and the first boundary condition relies on the idea that
the Universe is singularity-free, while the second one
ensures a physical behavior at infinite radius of curvature. 

\section{III. Spectrum of the Quantum Universe}

In order to study the previous eigenvalue problem ($\ref{H}$)
we take the following integral representation for the wavefunction
$\chi(R,\beta_{\pm},\phi)$ evaluated in a fixed space point 

\be
\chi(R,\beta_{\pm},\phi)=\int \theta_K(R)F_K(R,\beta_{\pm},\phi)dK.  
\ee

Therefore, performing an adiabatic approximation
($|\p_RF|\ll|\p_R\theta|$)
(allowed by the vanishing behavior that the spatial curvature has
near the singularity),
we obtain the following
reduced problems:

\be\label{eqR}%
\kappa\f d {dR}\left(\f 1 R \f{d\theta}{dR} \right) + \left(\kappa \f{K^2}{R^3}+\f{\mu^2}{R}+\f{\sigma^2}{R^2}-\epsilon\right)\theta=0,
\ee

\be\label{eqbeta}%
-(\p_+^2+\p_-^2+\f3{8\pi\kappa}\p_\phi^2)F+\f{R^6}{4\kappa^2 l^2_{in}}V(\beta_{\pm})F=K^2(R)F.
\ee 

Because of the presence of the scalar field,
near the cosmological singularity ($R\rightarrow0$),
the dynamical role of the potential
term in the equation (\ref{eqbeta}) can be neglected \cite{Ki92}
as soon as the following condition holds

\be\label{concla}%
R^3 \ll \mathcal{O}\left(l_P^2 l_{in}\sqrt{\f{<K2>}{|\bar{V}(\beta_{\pm})|}}\right),
\ee

where $l_P=\sqrt{G}$ denotes the Planck length and

\be
\bar {V}(\beta_{\pm})\equiv\int_{\Delta\beta^2}d^2\beta{V(\beta_{\pm})},
\ee

where, instead of ideal monochromatic solutions,
we deal with real wave packets which are flat
over the width $\Delta\beta\sim 1/\Delta k_\beta$ ($\Delta k_\beta$ being the standard deviation in the momenta space).

Hence, the eigenvalue problem ($\ref{H}$)
reduces to
a system of $\infty^3$ independent eigenvalue problem
(in each space point isomorphic to a Bianchi I model).
The solution of (\ref{eqbeta}) now reads
as plane waves
(we replace the variables $\beta_\pm$ with
the polar ones $r$ and $\alpha$, in order
to outline the vanishing behavior that
the wave function assumes at the infinite of the
($r,\; \alpha $)-plane):

\begin{multline}
F(r,\alpha,\phi)=e^{i\sqrt{8\pi\kappa/3}k_\phi\phi}\cdot\\
\cdot\sum_{m=0}^\infty J_m(k_\beta r)[A_m\cos(m\alpha)+B_m\sin(m\alpha)],
\end{multline}

$J_m$ being the first type Bessel function and $k^2_\beta$ the anisotropies eigenvalue: $K^2=k^2_\beta+k^2_\phi$.

To solve the equation (\ref{eqR}) we use a function of the form

\be
\theta(R)=\omega(R)\exp\left[-\f 1 {2l_P^2}\left(R+\f{\epsilon l^2_P}{16\pi}\right)^2\right],
\ee

so requiring the probability density to decay
with a Gaussian weight in the region where
the potential term is important. Therefore we obtain
the following equation for $\omega$:

\be\label{eqome}%
\kappa\left[R^2\omega''+R\omega'\sum_{i=0}^2a_i R^i\right]+\omega\sum_{j=0}^4b_j R^j=0,
\ee

where

\be
a_0=-1, \quad a_1=-\epsilon/8\pi, \quad a_2=-2/l_P^2,
\ee

\begin{align}
b_0=\kappa K^2, \quad b_1&=\f{\kappa\epsilon}{16\pi}+\sigma^2, \nonumber \\
b_2=\kappa\left(\f{\epsilon l_P}{16\pi}\right)^2+\mu^2, \quad & b_3=0, \quad b_4=\f{\kappa}{l_P^4}.
\end{align}

As result about our belief that the
quantum behavior of Universe is confined near
the classical singularity,
we have to investigate equation (\ref{eqome})
only in the limit $R\rightarrow0$. Thus we can neglect
the higher-order terms $\mathcal O(R^4)$;
then the solution of the above equation reads 

\be\label{serie}%
\omega(R)=\sum_{n=0}^\infty c_nR^{n+\gamma}, \quad \gamma=1-\sqrt{1-K^2}, \quad c_0\neq0,
\ee

whose coefficients obey the following recurrence relations
\begin{multline}\label{rr}%
c_n=-f(n,\gamma)\left\{\left[-\f \epsilon {8\pi}\left(n+\gamma-\f 3 2\right)+\f{\sigma^2}{8\pi l_P^2}\right]c_{n-1}\right.+\\
+\left.\left[-\f 2 {l_P^2}(n+\gamma-2)+\left(\f\epsilon {16\pi}\right)^2+\f{\mu^2}{8\pi l_P^2}\right]c_{n-2}\right\},
\end{multline}

with $f(n,\gamma)=((n+\gamma)(n+\gamma-2)+K^2)^{-1}$. 
Since we required the wavefunction to
decay at large $R$, where the potential term
becomes relevant, then, to avoid its  divergence
far from the Big-Bang, the series (\ref{serie})
must therefore terminate, i.e.
$\omega(R)$ must be polynomial  
in $R$. From the recurrence relations (\ref{rr})
this condition yields

\be\label{spee}%
\epsilon_{n,\gamma}=\f{\sigma^2}{l_P^2(n+\gamma-1/2)}
\ee

\be\label{mue}%
2(n+\gamma)=\left(\f{l_P\epsilon_{n,\gamma}}{16\pi}\right)^2+\f{\mu^2}{8\pi}.
\ee

The previous relations stand for the spectrum
of the Universe super-Hamiltonian and
of the quantum number
associated to the ultrarelativistic term.
It is to be emphasized that 
the ground state $n=0$ eigenvalue 

\be
\epsilon_{0,\gamma }=-\f{\sigma^2}{l_P^2(1/2-\gamma)}
\ee

for $\gamma<1/2$ is negative;
thus the real ground state, according to equation (\ref{mue}) for $\mu^2=0$, corresponds to $\gamma \sim 1.8\cdot 10^{-3}$, i.e. 

\be
\epsilon_0\simeq-\f{2\sigma^2}{l_P^2}
\, , 
\ee

is associated to a positive dust energy density.

\section{IV. Quasi-Classical Limit of the Model}

To complete the study of the eigenvalue problem,
let us now show that the $R$-dependence of the wave function
$\chi$ approaches a quasi-classical limit before the potential
becomes important. To this end we take the following
representation

\be
\theta=\exp\left[\f i \hbar\left(\Sigma_0+\f \hbar i\Sigma_1\right)\right]
\ee

where,
by the full quantum equation, we get
(at zero and first order in $\hbar $ respectively)
$\Sigma_0=\sqrt{\f{|\epsilon|}{\kappa}}R^{3/2}+\mathcal{O}(\sqrt R)$ and $\Sigma_1\sim\ln R$.
For sufficiently large values of $R$,
the logarithmic correction decays and the solution
approaches the quasi-classical limit,
even if the limit $\hbar\rightarrow0$
is not taken. A good estimation for the achievement
of such quasi-classical limit is obtained by requiring
a large value of $\Sigma_0$

\be
R^3\gg\mathcal{O}\left(\f{l_P^2}{|\epsilon|}\right).
\ee

Comparing the above relation with inequality (\ref{concla}), we
establish the constraint which ensures that the quasi-classical limit is reached before the potential term becomes important, explicitly

\be
l_{in}\gg \sqrt{\f{|\bar{V}(\beta_{\pm})|}{<K^2>}}
\f 1{|\epsilon|}.
\ee

This constraint states the degree of inhomogeneity
allowed by the self-consistency of our scheme
and the prediction of the spectrum obtained
neglecting the potential; as far as the ground state
eigenvalue $\epsilon _{0,\gamma}$ is concerned, the above relation correlates the
inhomogeneous scale to the Planck length, i.e. the cosmological
horizon associated to the quantum era of the Universe.
 
On the other hand, we observe that the ground state
probability density is a Gaussian distribution, peaked around
$R = \sigma ^2/8\pi$ by the width $l_P/\sqrt2$, i.e.

\be
|\theta_0(R)|^2 = \frac{1}{2\pi l_P^2}
\exp\left[ -\f 1 {l_P^2}\left(R -\f{\sigma^2}{8\pi}\right)^2\right] ; 
\label{Gaus}
\ee

such a distribution provides a singularity-free behavior of the
early Universe and, in the limit $l_P \rightarrow 0$, it
approaches,
according to the classical limit request, 
the Dirac delta-function
$\delta (R - \sigma ^2/8\pi)$.

Finally we face the question of the classical limit
of the spectrum in the
sense of large occupation numbers
$n\rightarrow \infty$. As the Universe approaches excited states,
the eigenvalue $\epsilon$ takes positive values and the
associated dust energy density would be negative.
However in the limit of very high occupation numbers,
the eigenvalue approaches zero as $1/n$ and no significant
phenomenology can come out from the negative energy density.
In other words, for very large $n$, our quantum dynamics
would overlap the Wheeler-DeWitt approach
(characterized just by $\epsilon \equiv 0$).

Summarizing we expect that for large enough values of
$R$, the quantum dynamics reaches a quasi-classical
representation and, regarding $\hbar $ as small, the
ground state wavefunction provides a localized initial
condition for the further evolution.

As a next step we want to investigate the
phenomenological implications of the obtained 
spectrum in view of the role $\epsilon$ can play
as dark matter candidate.

\section{V. Phenomenology of the dust fluid: 
discussion and conclusions}

In order to investigate the implications of the
spectrum (\ref{spee}) with respect to the formulation
of a dark matter candidate
via the non-zero eigenvalue, since the existence of a
Plackian lattice for the gravitational field is a well grounded
statement \cite{Rov,Pol}, then we put by hands a cut-off
length in our model. In the eigenvalues problem (\ref{H})
we deal with a perfect gas contribution and therefore
we have to implement a minimal length $l$ per particle,
that is (in this section we restore $\hbar$ and $c$)

\be
l^3 \equiv \f V {\cal{N}}=\f{3}{2}\f{\hbar cl_P}{\rho_{pg}\lambda^2},
\ee     

where $\cal{N}$ is the number of particles,
$\lambda$ the thermal length of the particles
and $\rho_{pg}$ the perfect gas energy density,
i.e. $\rho_{pg}=\sigma^2/R_P^5\le \sigma ^2/l_P^5$ ($R_P$ being the 
radius of curvature at the Planck era, to be regarded as the minimal one).
Thus, by requiring $l\geq l_P$, we get the following
inequality for the quantum number $\sigma^2$

\be\label{consigma}%
\sigma^2\leq\mathcal{O}\left(\hbar cl_P\right)         
\ee 

as soon as it is recognized that the thermal
length of a Planck mass particle can be suitably estimated
by the Planck one
(the quantum Universe has the Planck temperature
as an upper limit).
This constraint on the range of values available 
$\sigma ^2$ implies some relevant phenomenological issues.

\vspace{0.6cm}

i)---The upper limit for $\sigma ^2$ ensures that
the spectrum is limited by below and we have
$|\epsilon|\le \mathcal{O}(\hbar c/l_P) \sim \mathcal{O}(10^{19}GeV)$. 
Since in the limit $\hbar \rightarrow 0$ the quantity
$\sigma ^2$
vanishes like $\hbar ^{3/2}$, we see that the
ground state probability density behaves like
$\delta (R)$ and the singular behavior of the classical
Universe is restored. 

\vspace{0.6cm}

ii)---The contribution of our dark matter candidate 
to the actual critical parameter is provided by 

\be
\Omega_{dm}\sim
\f{\rho_{dm}}{\rho_{Today}c^2} \sim
\mathcal{O}\left(\f{\epsilon}{R^3_{Today}}\f{d_H^2 l_P^2}{\hbar c} \right)
\, ,
\ee

where $\rho_{Today} = 1.88h^2\times 10^{-29}g/cm^3$
($h\in[0.4,1]$) denoting the present critical density,
$R_{Today} \sim \mathcal{O}(10^{28}cm)$ the present radius
of curvature and
$d_H \sim \mathcal{O}(10^{27}cm)$ is the Hubble size
of the Universe.
Using the relation
$\epsilon _0 \sim \hbar c/l_P$,
the critical parameter $\Omega _{dm}$ rewrites

\be
\Omega _{dm} \sim \f{d_H^2 l_P}{R^3_{Today}}\sim
\mathcal{O}\left( 10^{-60}\right)
\label{dmn}
\ee

Such a parameter is much less than unity
and it can not be recognized as the dark matter
contribution 
(the dark matter critical parameter being estimated
$\mathcal{O}(0.3)$). 

Thus, our analysis has shown that the quantum
energy of the gravitational field, described by
an evolutionary theory, is not a suitable candidate
for {\it cold dark matter};
from a phenomenological point of view the evolutionary and
the Wheeler-DeWitt scheme overlap.

\vspace{0.6cm}

iii)---It is worth stressing that to
have, for our candidate, 
a critical parameter order of unity, 
it is necessary a ground state energy of
$\mathcal{O}(10^{60})$ times $\epsilon _0$ and,
as discussed in \cite{CoMo04} for the case of the
Friedmann-Robertson-Walker model,
this would simply corresponds to fix the Planck radius
of curvature $R_{P}$ about to $\mathcal{O}(10^{-21}cm)$,
instead of the Planck length taken above. 
Despite the idea of a Planckian Universe much larger than
the Planck volume is acceptable, in the context of our model,
two facts are against it.

a) The Planckian radius associated to the cut-off
$R_P^{cut}$, the classical one corresponding to
$\epsilon _0$ $R_P^{clas}$ and the mean value of $R$ in the
ground state $<R>$, respectively read

\be
R_P^{cut} \sim l_P\sqrt[5]{\frac{l_P\epsilon _0}{\hbar c}}
\, ,
R_P^{clas} \sim l_P\sqrt[3]{\frac{l_P\epsilon _0}{\hbar c}}
\, ,
<R> \sim l_P\left( \frac{l_P\epsilon _0}{\hbar c}\right);
\ee

these three values are of the same order ($l_P$),
as they must, only for
$\epsilon _0 \sim \hbar c/l_P$ (like above) and
when $\Omega _{dm} \sim 1$, i.e. for
$\epsilon _0\sim \mathcal{O}(10^{60}\hbar c/l_P)$,
they deeply differ.

b) For the inflationary scenario to take place in its standard
formulation (at the energy scale $\Sigma\sim\mathcal{O}(10^{14}GeV)$), we have
to require the dust energy density to be smaller than the
vacuum energy and this leads to the following value for the
radius of curvature $R_i$ when inflation starts

\be
R_i = \left( \frac{\epsilon _0}{\Sigma }\right) ^{4/3}
\frac{\hbar c}{\epsilon _0}
\, ;
\ee 

If $\epsilon _0\sim \hbar c/l_P$, then inflation starts, as
in the standard case, at
$R_i\sim \mathcal{O}(10^6)l_P$, while for the value of
$\epsilon _0\sim \mathcal{O}(10^{60}\hbar c/l_P)$ we get
an initial radius of curvature $20$ orders of magnitude greater.
In this second case inflation would start
when the Universe has a radius of curvature too large for
implementing a suitable de-Sitter phase; in fact, the
corresponding \emph{e-folding} would be here
about $23$ and this value would be not enough to solve
the horizon paradox.

\vspace{0.6cm}

Thus, we can conclude that, from a phenomenological point of
view, the Evolutionary Quantum Cosmology overlaps, in the
generic inhomogeneous case, the Wheeler-Dewitt description
and no evidence appears of the non-zero eigenvalue in the
Universe critical parameter. As a final point, we want to emphasize how the Planckian
dimension of the quantum Universe, predicted by our model,
provides good indications on the solution of the horizon
paradox within a Quantum Mechanics endowed with a cut-off.
In fact, if (like here) the mean size of the primordial Universe
is comparable to the classical horizon at the Planck time,
no real puzzle arises about its later strong uniformity.

\section{AKNOWLEDGMENTS}
We would like to thank Alessandra Corsi 
for her valuable comments on this topic and the
help to improve its presentation. We thank the Referee for his 
clever comments on this topic which allowed us to improve and complete our analysis.

\end{document}